\documentclass[final]{siamltex1213}
\usepackage{amsmath}
\usepackage{amsfonts}
\usepackage{graphicx,times,bm,url}
\usepackage{afterpage,lscape}
\usepackage{rotating}
\graphicspath{{./fig/}{./png/}}

\newcommand{\EQ}{\begin{equation}}
\newcommand{\EN}{\end{equation}}
\newcommand{\EQA}{\begin{eqnarray}}
\newcommand{\ENA}{\end{eqnarray}}

\newcommand{\Fig}[1]{Figure~\ref{#1}}

\newcommand{\Tabs}[2]{Tables~\ref{#1} and \ref{#2}}

{}
{}
{}

{}
{}
{}
{}
{}
{}
{}
{}
{}
{}
{}
{}
{}
{}
{}
{}
{}
{}
{}
{}
{}

{}
{}
{}

{}

{}
{}

\newcommand{\eee}{\hat{\mbox{\boldmath $e$}} {}}
\newcommand{\nnn}{\hat{\mbox{\boldmath $n$}} {}}

\newcommand{\rr}{\mbox{\boldmath $r$} {}}

\newcommand{\xx}{\bm{x}}
\newcommand{\XX}{\bm{X}}

\newcommand{\BB}{\bm{B}}

\newcommand{\uu}{\mbox{\boldmath $u$} {}}

\newcommand{\JJ}{\mbox{\boldmath $J$} {}}

\newcommand{\nn}{\mbox{\boldmath $n$} {}}

\newcommand{\nab}{\mbox{\boldmath $\nabla$} {}}

\newcommand{\dd}{{\rm d} {}}

\def\EM{E_{\rm M}}

\def\EM{E_{\rm M}}

\title{Mimetic Methods for Lagrangian Relaxation of Magnetic Fields
\thanks{We acknowledge the use of the computing facilities HECToR,
part of the UK National Supercomputing Service in Edinburgh.
All the authors acknowledge financial support from the UK's
STFC (grant number STK/K000993/1).
We gratefully acknowledge the support of NVIDIA Corporation with the
donation of one Tesla K40 GPU used for this research.
We are grateful for fruitful discussions with Antonia Wilmot-Smith,
Irene Kyza and Ping Lin.
We greatly appreciate the comments and suggestions by the anonymous
referees.
}}

\author{
S. Candelaresi\footnotemark[2]\and
D. Pontin\footnotemark[2]\and
G. Hornig\footnotemark[2]
}

\begin{document}
\maketitle

\renewcommand{\thefootnote}{\fnsymbol{footnote}}
\footnotetext[2]{Division of Mathematics, University of Dundee, Dundee, UK
(simon.candelaresi@gmail.com)}
\renewcommand{\thefootnote}{\arabic{footnote}}

\slugger{sisc}{xxxx}{xx}{x}{x--x}

\begin{abstract}
We present a new code that performs a relaxation of a magnetic field towards
a force-free state (Beltrami field) using a Lagrangian numerical scheme.
Beltrami fields are of interest for the dynamics of many technical and
astrophysical plasmas as they are the lowest energy states that the magnetic
field can reach.
The numerical method strictly preserves the magnetic flux and the topology of magnetic
field lines.
In contrast to other implementations we use mimetic operators for the
spatial derivatives in order to improve accuracy for high distortions
of the grid.
Compared with schemes using direct derivatives we find
that the final state of the simulation approximates a
force-free state with a significantly higher accuracy.
We implement the scheme in a code which runs on graphical processing units
(GPU), which leads to an enhanced computing speed compared to previous
relaxation codes.
\end{abstract}



\pagestyle{myheadings}
\thispagestyle{plain}
\markboth{S. CANDELARESI, D. PONTIN, G. HORNIG}
{MIMETIC METHODS FOR MAGNETIC FIELD RELAXATION}

\section{Introduction} 

For astrophysical plasmas magnetic diffusivity
can be so low that one can assume the plasma to evolve on dynamic
timescales according to the ideal induction equation
\begin{equation} \label{eq: ideal induction}
\frac{\partial {\BB}}{\partial t} - \nab\times(\uu\times\BB)=\bm{0},
\end{equation}   
where $\BB$ is the magnetic field and $\uu$ the plasma velocity.
Such an evolution equation is most conveniently studied using the Lagrangian
description of the fluid, with position vectors of fluid elements represented by $\xx(\XX,t)$,
where $\xx(\XX,0) = \XX$.
For our purposes we assume that $\xx(\XX,t)$ is differentiable for all $\XX$ and $t$.
Equation \eqref{eq: ideal induction}
implies that magnetic field lines behave like material
lines of the plasma \cite{alfven1943,BatchelorFrozeIn1950RSPSA,PriestReconnection2000},
a property which can be expressed with the help
of the flow of the velocity field, 
\begin{equation}
\frac{\partial \xx(\XX,t)}{\partial t} = \uu\left(\xx(\XX,t),t\right),  \label{eq: x integration}
\end{equation} 
which together with equation \eqref{eq: ideal induction} implies that
the magnetic fields at time $t=0$ and at $t>0$ are related by the pull-back
under $\xx(\XX,t)$ (see appendix \ref{sec: induction lie}):
\begin{equation}
(\xx^{*}(t)\BB)(\XX,t) = \BB(\XX,0). \label{eq: pull back}
\end{equation}
This is a modern formulation of Alfv{\' e}n's Theorem \cite{alfven1943}.
Here and in the rest of this paper we express the magnetic field $\BB(\XX,t)$
as a function of the initial grid positions $\XX$ and time $t$.
We can also express it as function of the coordinates with the different
functional form $\tilde{\BB}(\xx(\XX,t),t) = \BB(\XX,t)$.
When describing our numerical scheme we will sometimes for simplicity suppress
the explicit time dependence and simply write $\BB(\XX,t) = \BB(\XX)$.

The velocity in Equation (\ref{eq: ideal induction}) is coupled to the magnetic field via the
magnetohydrodynamic (MHD) momentum balance equation in a highly non-linear way. 
However, in order to determine the end state of such an evolution (in the absence of 
external forces) one only has to know that the presence of a non-zero viscosity will continuously
extract energy from the system until a minimum energy state is reached \cite{Moffatt1985}. 
In the absence of significant gas pressure (low plasma-$\beta$) this minimum energy
state can be obtained from a simple variation of the magnetic energy density
under the assumption of an ideal evolution (Eq.~\ref{eq: ideal induction}).
This results in a condition for a so-called force-free field or Beltrami field
\EQ
(\nab\times\BB) \times\BB = 0 \ \Leftrightarrow \
\nab\times\BB = \alpha\BB.
\EN
Here $\alpha$ is in general a function of the spatial variables, but due to
the solenoidal condition on $\BB$, $\alpha$ has to be constant along field
lines, i.e.\ $\BB\cdot\nab \alpha = 0$.    

As long as one is interested only in the minimum energy state of the evolution,
one can also prescribe an artificial dynamics instead of using the MHD momentum 
balance equation.
Specifically, if one takes 
 \begin{equation}
\uu = \gamma\JJ\times\BB; \quad \JJ = \nab\times\BB; \quad \gamma > 0, \label{eq: magneto friction}
\end{equation}  
it is easy to prove that the magnetic energy monotonically decreases until a
force-free state is reached
\cite{Craig-Sneyd-1986-311-451-ApJ,Yang-Sturrock-1986-309-383-ApJ}.
This approach is called the magneto-frictional method \cite{Chodura1981}. 
$\JJ$ is the electric current density, where we normalize by setting the permeability $\mu_0=1$.

The question as to whether, for an arbitrary given initial field $\BB(\XX, 0)$,
a corresponding Beltrami field with the same topology
(i.e.~satisfying \eqref{eq: pull back} for some mapping $\xx$)
exists, and if so whether it is smooth, is unsolved.
Examples where the corresponding Beltrami fields have singularities
(typically current sheets) exist \cite{Syrovatskii71}.
These weak solutions occur in particular for cases where points, lines or
surfaces of vanishing magnetic field strength exist in the initial field.
A debate is still ongoing under which conditions non-smooth solutions can develop
from smooth initial fields in regions of non-vanishing magnetic field 
 \cite{Craig2005,longcope1994,low2010apj,Parker72,rapazzo2013,vanballegooijen1985,wilmotsmith2009a}.
This question was first raised by E.\ Parker, as a possible scenario for
the onset of magnetic reconnection in the solar atmosphere, and is also known
as the Parker Problem.   

Studying magneto-frictional relaxation numerically with an Eulerian
description requires high spatial resolution in order to reduce
numerical diffusion.
However, the numerical diffusion can never be completely eliminated with such
a standard approach.
Consequently, an ideal evolution preserving the topology of $\BB$ can only
be approximated.
In order to circumvent this problem, Craig et al.\ \cite{Craig-Sneyd-1986-311-451-ApJ}
used a Lagrangian approach which directly calculates the mapping $\xx(\XX,t)$ and 
hence simulates a perfectly ideal evolution.  
The method therefore preserves the topology of
field lines as well as the magnetic flux through each surface element.
Additionally,
$\nab\cdot\BB = 0$ is automatically preserved, thus eliminating
the need for divergence cleaning.

Later, Pontin et al.\ \cite{Pontin-Hornig-2009-700-2-ApJ}
analyzed the quality of the force-free approximation obtained using the
method of Craig et al.\ \cite{Craig-Sneyd-1986-311-451-ApJ}.
They found that, while the numerically calculated value of
$(\nab\times\BB)\times \BB$ could be minimized to an arbitrarily small value,
the true value of $(\nab\times\BB)\times \BB$---obtained from independent
measures described below---was sometimes much higher.
The reason for this discrepancy was identified to be numerical errors in
the derivatives that increase as the grid becomes highly distorted.
Accumulation of these errors occurs due to several
multiplications of first and second derivatives that are required to obtain
an expression for $(\nab \times \BB) \times \BB$ in the scheme
(see Eqs.~(2.11) and (2.12) of Craig et al.\ \cite{Craig-Sneyd-1986-311-451-ApJ}).
Consequently, it turns out that $\nab\cdot\JJ \ne 0$ can become large for
high grid distortions.
It was suggested by Pontin et al.\ \cite{Pontin-Hornig-2009-700-2-ApJ} that these errors
could be reduced by calculating the electric current density using
a mimetic method
\cite{Hyman-Shashkov-1997-33-4-CompAthApp, Hyman-Shashkov-1999-151-2-JCompPh}.
Derivative operators are then represented as integrals, by making use of
e.g.\ Stokes' or Gauss' theorem.
One of the great advantages of this approach is that numerically computed
curls are discretely divergence free.

In the present work we apply these methods with the two-fold aim of a qualitative
improvement of the force-free approximation obtained and a faster convergence.
In order to assess the quality of the force-free approximation and computational
efficiency of our new scheme, we also implement the classical method, as described
by Craig et al.\ \cite{Craig-Sneyd-1986-311-451-ApJ}.
The two methods are compared throughout the remainder of the paper.

\section{Numerical Approach: The GLEMuR Code}

\subsection{Magnetic Field Relaxation}

For the evolution of the velocity field $\uu$ we use the aforementioned
magneto-frictional force \eqref{eq: magneto friction}, as it causes the
magnetic energy to decay monotonically and the field to evolve towards
a force-free state.
For the sake of simplicity the parameter $\gamma = 1$ is chosen to be constant
in time and space.
In principal it can depend on space and time and this can be used 
for instance to address concerns about the magneto-frictional method raised by  
Low \cite{Low-2013-768-7-ApJ} for cases of fixed boundaries or null
points in the domain.
All examples discussed below,  however, do not require this. 

From the pull-back formula, \eqref{eq: pull back}, an equation for the magnetic
field can be derived \cite{MoffattBook1978} (see appendix \ref{sec: induction lie})
\begin{eqnarray} \label{eq: B transformation}
B_{i}(\XX,t) &=& \frac{1}{\Delta} \sum_{j=1}^3\frac{\partial x_{i}}{\partial X_{j}} B_{j} (\XX,0), 
\label{eq: B evolution}
\end{eqnarray}
where $\Delta$ is the determinant of the Jacobian matrix
$\partial x_{i}/\partial X_{j}$ and measures the local compression or expansion
of the medium.
This is analogous to Nanson's formula known in continuum mechanics.
Equation \eqref{eq: B evolution} is used to determine $\BB$ in the 
Lorentz force, which is required for the numerical integration of \eqref{eq: x integration}. 
The other quantity required for the Lorentz force is the electric current which we 
determine from $\BB$ via a mimetic operator.  

\subsection{Mimetic Operators}\label{sec: nearest}

A property of the mimetic differential operators described by Hyman et al.\
\cite{Hyman-Shashkov-1997-33-4-CompAthApp} is that they map fields defined on
a discrete space, like grid points,
onto a different discrete space, e.g.\ centers of grid faces.
The curl operator maps the magnetic field, defined on grid nodes (primal mesh),
onto points in the centers of the faces of grid cells (dual mesh),
with the result that $\BB$ and $\JJ$ are known at
different locations.
This is a general characteristic of mimetic operators, which
map their result
onto edges, faces or cells, rather than onto the same grid points.

Terms like $(\nab\times\BB)\times\BB$ require $\BB$ and $\nab\times\BB$ 
to be known at the same locations.
Therefore, using the standard mimetic operators
some sort of interpolation needs to be applied.
It is not obvious which method or order of interpolation leads to numerical
accuracy or stability.
Although we do not have a mathematical proof on numerical stability,
we will characterize situations for which interpolations may fail.

Here we take an alternative approach, as described by Pontin et al.\
\cite{Pontin-Hornig-2009-700-2-ApJ},
that mitigates this requirement for an explicit interpolation step. 
The current through a surface $U$ bounded by the closed loop $C$ can in general be computed using Stokes' theorem:
\EQ \label{eq: Stokes integral}
\int\limits_{U}\JJ\cdot\nnn \ {\rm d}S =
\oint\limits_{C} \BB\cdot{\rm d}\rr .
\EN
\begin{figure}[t!]\begin{center}
\includegraphics[width=0.7\columnwidth]{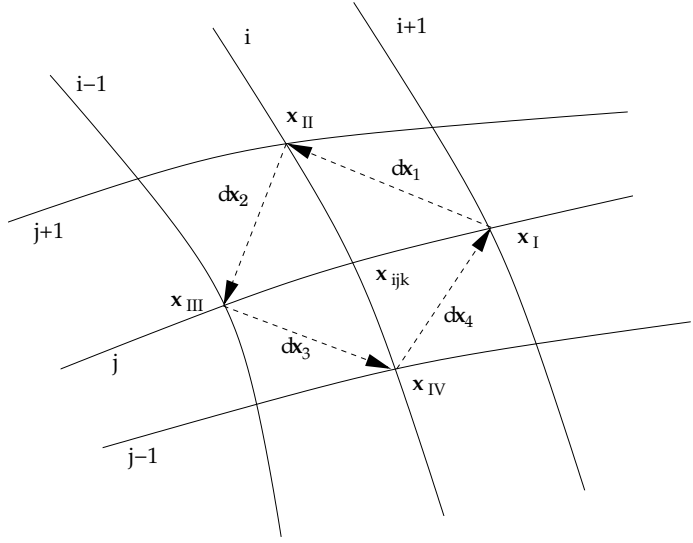}
\end{center}
\caption[]{
Schematic representation of the vectors used for the calculations 
in equations \eqref{eq: current discrete}--\eqref{eq: sum of triangles}.
Here only the contribution from the $ij$-index plane is shown.
For the remaining components one simply needs to cyclically permute the indices
$i$, $j$ and $k$.
}
\label{fig: stokesSchematic}\end{figure}
For the current at the grid point $\xx_{ijk} = \xx(\XX_{ijk},t)$ we calculate three loop integrals 
in the three grid surfaces which intersect at this point.
For the $ij$-grid surface this loop 
is shown in \Fig{fig: stokesSchematic}.
The right hand side of \eqref{eq: Stokes integral} is evaluated as 
\EQ \label{eq: current discrete}
  \sum_{r=1}^{4} \BB_{r}\cdot{\rm d}\xx_{r},
\EN
with the difference vectors ${\rm d}\xx_{i}$ defined as
${\rm d}\xx_{1} = \xx_{\rm II} - \xx_{\rm I}$,
${\rm d}\xx_{2} = \xx_{\rm III} - \xx_{\rm II}$, etc., the magnetic field
$\BB_{1} = (\BB(\XX_{\rm I}) + \BB(\XX_{\rm II}))/2$, etc.\ and the
position vectors $\xx_{\rm I} = \xx(\XX_{\rm I})$, etc.,
where we use the short hand notation $\xx(\XX) = \xx(\XX,t)$.
The left hand side of \eqref{eq: Stokes integral} we approximate by assuming that the current 
is constant on the quadrilateral,
\EQ
 \JJ(\XX_{ijk}) \cdot \sum_{r=1}^{4}\nn_{r} A_{r},
\EN
with the four triangle elements
\begin{eqnarray} \label{eq: four surfaces}
\nn_{1}A_{1} & = & (\xx_{\rm I}-  \xx_{ijk})\times(\xx_{\rm II}- \xx_{ijk})/2, \nonumber \\
\nn_{2}A_{2} & = & (\xx_{\rm II}- \xx_{ijk})\times(\xx_{\rm III}-\xx_{ijk})/2, \nonumber \\
\nn_{3}A_{3} & = & (\xx_{\rm III}-\xx_{ijk})\times(\xx_{\rm IV}- \xx_{ijk})/2, \nonumber \\
\nn_{4}A_{4} & = & (\xx_{\rm IV}- \xx_{ijk})\times(\xx_{\rm I}-  \xx_{ijk})/2.
\end{eqnarray}
The sum of the four surface elements is given as
\begin{eqnarray} \label{eq: sum of triangles}
\nn A = & & \sum_{r=1}^{4}\nn_{r}A_{r} 
= \left({\rm d}\xx_{1}\times{\rm d}\xx_{2} + {\rm d}\xx_{2}\times{\rm d}\xx_{3}
  + {\rm d}\xx_{3}\times{\rm d}\xx_{4} + {\rm d}\xx_{4}\times{\rm d}\xx_{1}\right)/4.
\end{eqnarray}
Hence, the discretized version of \eqref{eq: Stokes integral} at the grid location
$\xx_{ijk}$ reads
\EQ \label{eq: current matrix equation}
\JJ(\XX_{ijk})\cdot\nn= \frac{1}{A} \sum_{r=1}^{4} \BB_{r}\cdot{\rm d}\xx_{r}.
\EN
This equation determines the current in direction $\nn$ at $\xx_{ijk}$.
Together with 
corresponding loops in the $jk$- and $ki$-grid surfaces complete information of all 
three components of the vector $\JJ$ is obtained. 
The current components in the $X,Y$ and $Z$ directions are related to the projections of $\JJ$ 
along $\nn$ by a system of linear equations that can be solved by inverting the matrix
composed of the three normal vectors $\nn$.
Note that the three normal vectors 
have to be linearly independent, which will always be the case so long as the grid does 
not collapse to being locally two-dimensional. 
We note that the scheme provided above makes use of a modified version of the 
approximated normal compared to that used by
Pontin et al.\ \cite{Pontin-Hornig-2009-700-2-ApJ}.

While the above approach removes the requirement of an explicit interpolation step,
we note that it is based on the assumption that $\JJ$ can be approximated as
constant over the quadrilateral shown in \Fig{fig: stokesSchematic}.
For distortions on the grid scale, e.g.\ foldings,
this approximation will no longer be appropriate.

\subsection{Next Nearest Neighbor Mimetic Approach}\label{nextnearestsec}

For finite difference methods, a higher order scheme, including further
next nearest neighbors, may increase the stability and accuracy of the
numerical simulation\footnote{Higher order derivatives do not necessarily
lead to higher accuracies.
For sufficiently smooth solutions, however, they
lead to increased stability and accuracy for most practical problems.}.
To test whether accuracy and stability increase with a modified loop
integral for $\JJ$ we perform a similar calculation as in equation
\eqref{eq: current matrix equation} using values like $\xx_{i+1,j+1,k}$
and the magnetic field on those grid points.

\begin{figure}[t!]\begin{center}
\includegraphics[width=0.7\columnwidth]{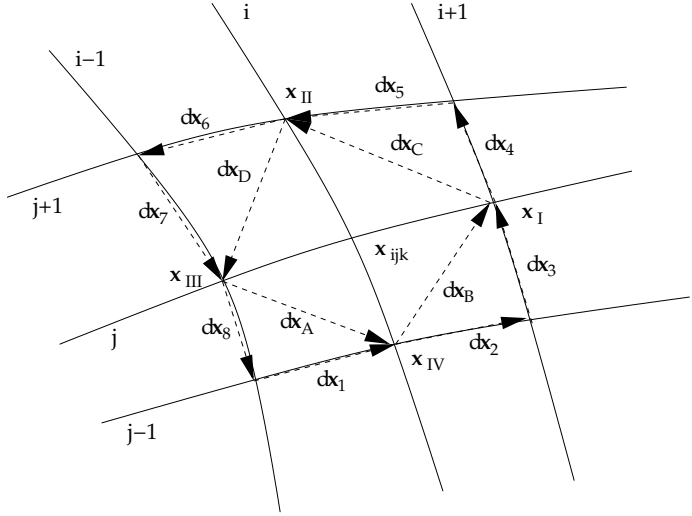}
\end{center}\caption[]{
Schematic representation of the vectors used for the calculations 
in equations \eqref{eq: eight surfaces} and
\eqref{eq: sum of eight triangles}.
Here only the contribution from the $ij$-index plane is shown.
For the remaining components one simply needs to cyclically permute the indices
$i$, $j$ and $k$.
}\label{fig: stokesSchematic4th}\end{figure}

Equation \eqref{eq: current discrete} is here augmented by further neighbors
of the point $\xx_{ijk}$, forming an octilateral (\Fig{fig: stokesSchematic4th}).
The loop integral over this octilateral includes eight contributions:
\EQ
\sum_{r=1}^{8}\BB_{r}\cdot{\rm d}\xx_{r},
\EN
where the difference vectors ${\rm d}\xx_{r}$ and the magnetic field vectors $\BB_{r}$
are chosen in analogy to equation \eqref{eq: current discrete}
(see \Fig{fig: stokesSchematic4th}).
The surface elements are
\begin{eqnarray} \label{eq: eight surfaces}
& \nn_{1}A_{1} = {\rm d}\xx_{2}\times{\rm d}\xx_{3}/2, \quad
\nn_{2}A_{2} = {\rm d}\xx_{4}\times{\rm d}\xx_{5}/2, \nonumber \\
& \nn_{3}A_{3} = {\rm d}\xx_{6}\times{\rm d}\xx_{7}/2, \quad
\nn_{4}A_{4} = {\rm d}\xx_{8}\times{\rm d}\xx_{1}/2, \nonumber \\
& \nn_{5}A_{5} = {\rm d}\xx_{A}\times{\rm d}\xx_{B}/4, \quad
\nn_{6}A_{6} = {\rm d}\xx_{B}\times{\rm d}\xx_{C}/4, \nonumber \\
& \nn_{7}A_{7} = {\rm d}\xx_{C}\times{\rm d}\xx_{D}/4, \quad
\nn_{8}A_{8} = {\rm d}\xx_{D}\times{\rm d}\xx_{A}/4.
\end{eqnarray}
The sum of the surface elements results in a similar equation as
\eqref{eq: sum of triangles}:
\begin{eqnarray} \label{eq: sum of eight triangles}
& \nn A = \sum\limits_{r=1}^{8}\nn_{r}A_{r} =
 \left({\rm d}\xx_{2}\times{\rm d}\xx_{3} + {\rm d}\xx_{4}\times{\rm d}\xx_{5} + 
 {\rm d}\xx_{6}\times{\rm d}\xx_{7} + {\rm d}\xx_{8}\times{\rm d}\xx_{1}\right)/2 \nonumber \\
& + \left({\rm d}\xx_{A}\times{\rm d}\xx_{B} + {\rm d}\xx_{B}\times{\rm d}\xx_{C}
  + {\rm d}\xx_{C}\times{\rm d}\xx_{D} + {\rm d}\xx_{D}\times{\rm d}\xx_{A}\right)/4.
\end{eqnarray}
Those elements are used in equation \eqref{eq: current matrix equation}, where
the matrix is inverted to calculate $\JJ(\XX_{ijk})$.

\subsection{Time Stepping}

For the numerical integration of equation \eqref{eq: magneto friction},
we are interested in fast convergence and stability.
Adaptive time steps are needed to keep the error within limits.
Therefore, we use the method of lines to express the partial differential
equations as a set of ordinary differential equations and
apply the fifth-order adaptive time step Runge--Kutta formula
\cite{Cash-Karp-1990-16-3-ACMTransMathSoft, NumericalRecipes3} for the
time stepping.

The time step is adjusted according to the error of the calculation.
If the error in $\xx$ exceeds a prescribed limit the step length is reduced via
\EQ \label{eq: time stepping}
{\rm d}t' = {\rm d}t\left| \frac{\Lambda_{0}}{\Lambda} \right|^{0.2},
\EN
where ${\rm d}t$ and ${\rm d}t'$ are the old and adjusted time steps,
$\Lambda$ the maximum error in $\xx$, as calculated in \cite{NumericalRecipes3}
and $\Lambda_{0}$ the desired maximum error (tolerance).
If $\Lambda > \Lambda_{0}$ the result is rejected and recomputed with
${\rm d}t$ = ${\rm d}t'$.
Should $\Lambda$ fall below $\Lambda_{0}/2$, ${\rm d}t'$ is increased
according to the same equation, thus accelerating computation.

As we are dealing with a highly parallelizable problem, we make use of
parallel computing facilities.
For that, we developed a numerical code named {\textsc GLEMuR}
(Gpu-based Lagrangian mimEtic Magnetic Relaxation) which
makes use of the computing power of graphical processing units.
As API we use CUDA \cite{Nickolls-Buck-2008-6-2-Queue}, which has been
tested and has seen various applications in computational analysis.

\subsection{Boundary Conditions}

In the code we implement both periodic boundary conditions and so-called
line-tied boundary conditions.
A line-tied boundary is a boundary at which the plasma velocity is zero and
the magnetic flux through any surface element is fixed (i.e.\ $\BB\cdot\nn$ fixed).
Periodic boundaries for a moving grid need careful treatment,
since periodic grid positions would not be physically consistent.
In order to be consistent with equation \eqref{eq: B evolution}, for a periodic boundary
in, say, the $z$-direction we choose
\begin{eqnarray}
x_{i,j,f-1} & = & x_{ijl} \nonumber \\
y_{i,j,f-1} & = & y_{ijl} \nonumber \\
z_{i,j,f-1} & = & z_{ijl} - L_{z} - {\rm d}Z
\end{eqnarray}
for the lower boundary, where $f$ and $l$ are the indices for the
first and last inner points of the domain in the $z$-direction and ${\rm d}Z$
is the initial grid spacing in the $z$-direction.
By analogy, the upper boundary is set to
\begin{eqnarray}
x_{i,j,l+1} & = & x_{ijf} \nonumber \\
y_{i,j,l+1} & = & y_{ijf} \nonumber \\
z_{i,j,l+1} & = & z_{ijf} + L_{z} + {\rm d}Z.
\end{eqnarray}
With these conditions the magnetic field is automatically periodic, i.e.
\begin{eqnarray}
\BB(\XX_{i,j,f-1},t) & = & \BB(\XX_{ijl},t) \nonumber \\
\BB(\XX_{i,j,l+1},t) & = & \BB(\XX_{ijf},t).
\end{eqnarray}
In the results described in the following sections all boundaries are line-tied,
though periodic boundaries do not qualitatively affect these results.

\section{Field Relaxation}

\subsection{Initial Configuration}

\begin{figure}[t!]\begin{center}
\includegraphics[width=.35\columnwidth]{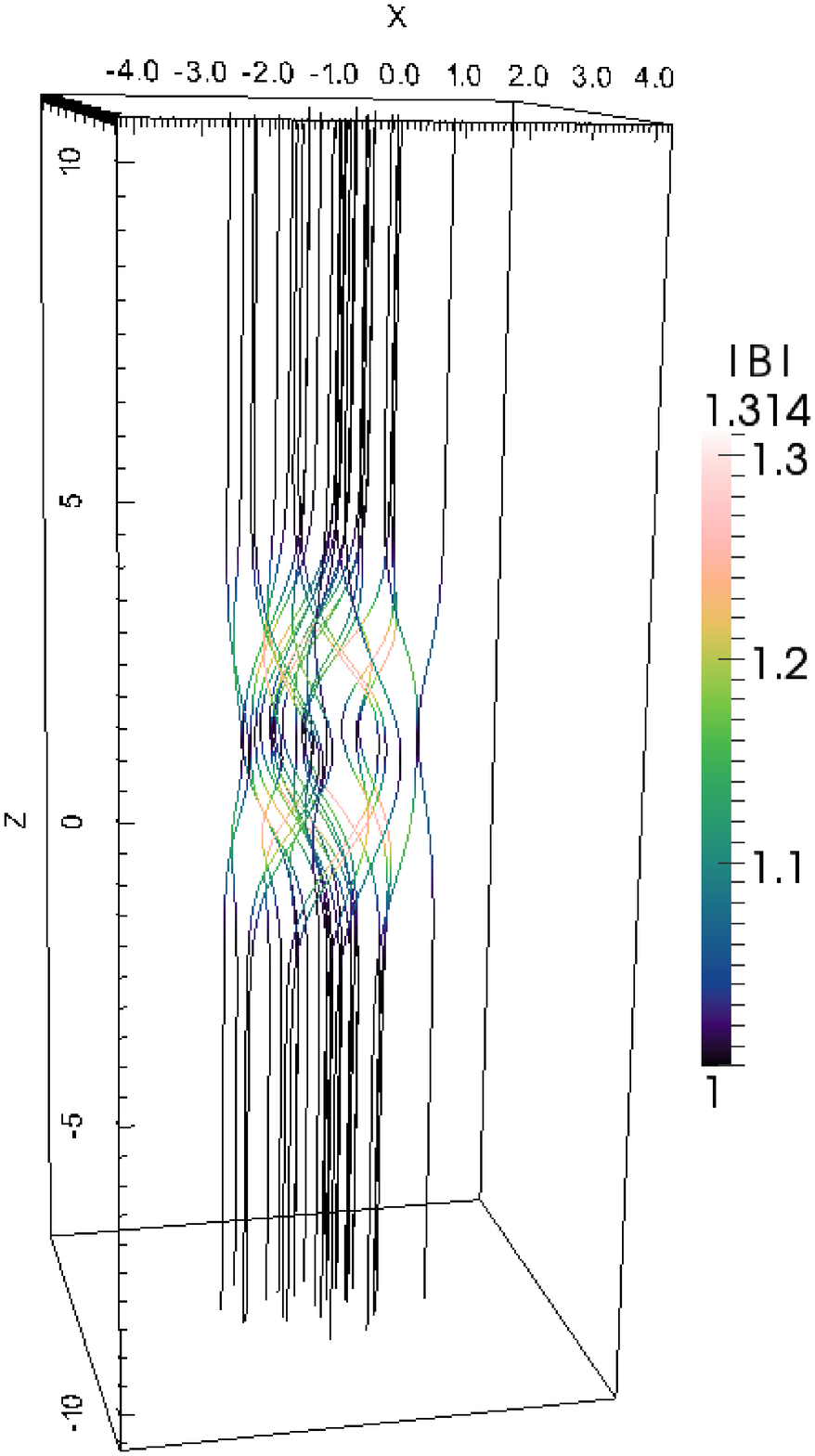}
\includegraphics[width=.35\columnwidth]{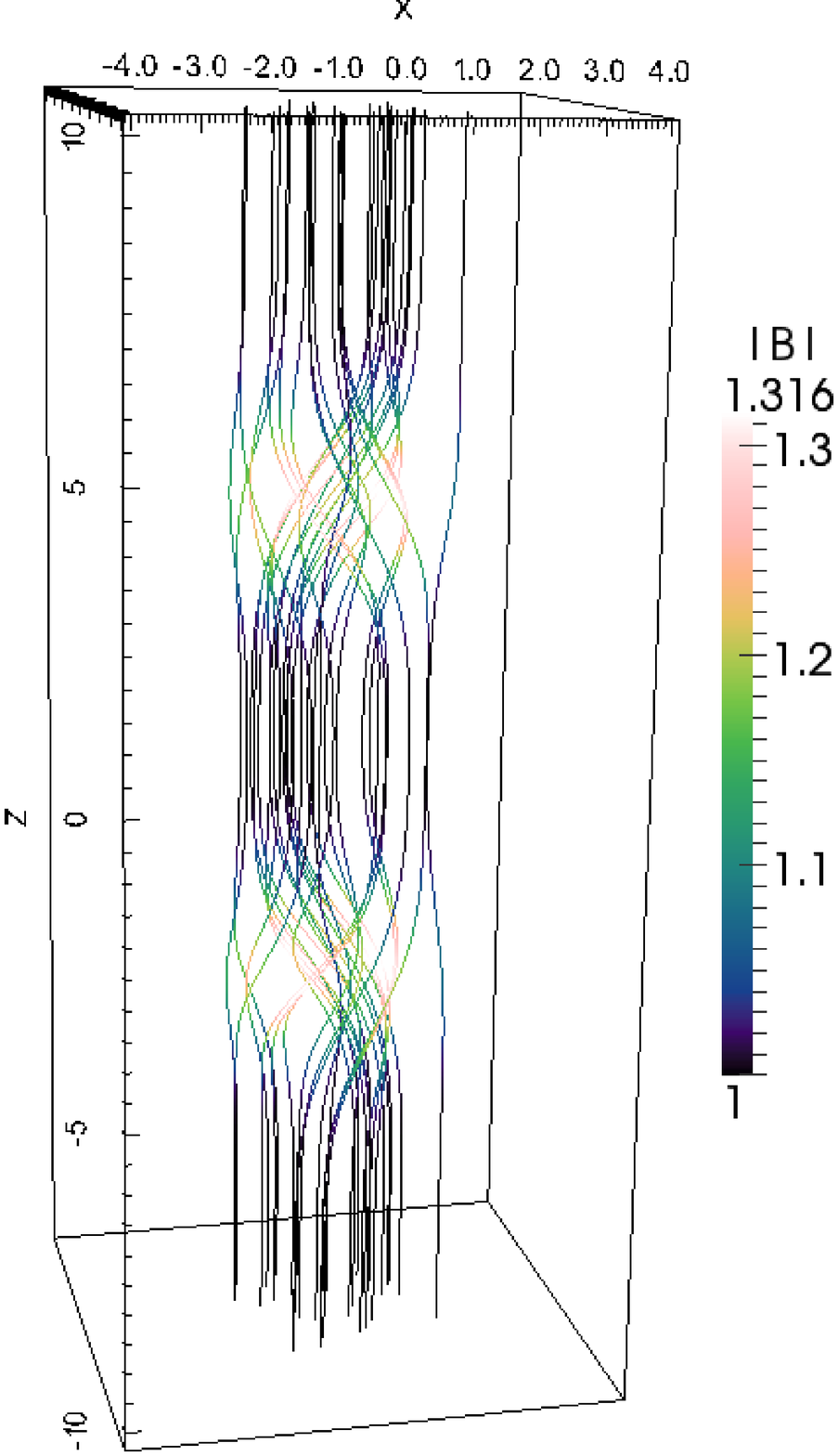}
\end{center}\caption[]{
Initial magnetic field for the {\em IsoHelix} field (left panel)
and the {\em Pontin09} field (right panel) with $\phi = \pi$ and
$\phi_{1/2} = \pm \pi$, respectively.
The colors denote the field strength.
For readability, only magnetic field lines passing the origin at a radius
of $2$ are plotted.
}\label{fig: B_init}\end{figure}

Using the GLEMuR code as described above
we compute the ideal evolution of initially twisted magnetic fields starting with a
rectangular computational grid.
For comparison purposes two initial magnetic field configurations are considered.
Our primary focus is on an initial field for which we have an exact closed-form
expression for the corresponding force-free field, i.e.\
we know exactly the expected values of $\xx(\XX, t \to \infty)$
and $\BB(\XX, t \to \infty)$. 
This allows us to compare in a straightforward and precise way the quality of the
relaxation.
The form of the initial magnetic field is given by
\EQA \label{eq: B init analytical}
\BB(\XX,0) & = & \frac{2B_{0}Z}{a_{z}^{2}}
 \exp\left(-\frac{X^{2}+Y^{2}}{a_{r}^{2}}-\frac{Z^{2}}{a_{z}^{2}}\right)\phi
 \left(Y\eee_{x}-X\eee_{y}\right)
 + B_{0}\eee_{z},
\ENA
with the initial magnetic field amplitude $B_{0}$,
length of the twist region $a_{z}$,
width of the twist region $a_{r}$, 
twist angle $\phi$ and
Cartesian unit vectors $\eee_{i}$.
Unless explicitly stated, we  choose $B_{0} = 1$, $a_{r} = \sqrt{2}$,
and $a_{z} = 2$.
The twist angle is chosen either $\phi = \pi/4$,
$\phi = \pi/2$ or $\phi = \pi$.
The domain is a cuboid with size $L_{x} = L_{y} = 8$ and $L_{z} = 20$ with
its center coinciding with the origin of the coordinate system
(\Fig{fig: B_init}, left panel).
Since field lines turn first by some angle around the $z$-axis and then back by 
the same angle, determined by the  $\phi$ parameter,
we will call this configuration {\em IsoHelix}.
Although the twist decreases like a Gaussian with distance to the center,
there is a small and negligible normal component at the side boundaries
of the order of $4.1\cdot10^{-5}$.

The expected magnetic field in the relaxed state is of the form
\EQ \label{eq: B relaxed}
\BB_{\rm relax} = B_{0}\eee_{z}.
\EN
For the same configuration we can compute the grid's deformation for
$t \to \infty$ which takes the form
\begin{eqnarray} \label{eq: x analytical relaxed}
\xx_{\rm relax} & = & \cos{\left(\exp{\left(-\frac{X^{2}+Y^{2}}{a_{r}^{2}}-\frac{Z^{2}}{a_{z}^{2}}\right)}\phi\right)}
                     (X\eee_{x}+Y\eee_{y}) \nonumber \\
                & &  + \sin{\left(\exp{\left(-\frac{X^{2}+Y^{2}}{a_{r}^{2}}-\frac{Z^{2}}{a_{z}^{2}}\right)}\phi\right)}
                     (Y\eee_{x}-X\eee_{y}) \nonumber \\
                & &  + Z\eee_{z}.
\end{eqnarray}

In the following we also mention results obtained using the identical initial
condition to that used by Pontin et al.\ \cite{Pontin-Hornig-2009-700-2-ApJ}.
They applied an initial magnetic field of the form
\EQA \label{eq: B init Pontin09}
\BB(\XX,0) &= & B_{0}\eee_{z} \nonumber \\
 & & + \sum_{i=1}^{2} \frac{2B_{0}\phi_{i}}{\pi a_{r}}
\exp\left(-\frac{X^{2}+Y^{2}}{a_{r}^{2}} - \frac{(Z-L_{i})^{2}}{a_{z}^{2}}\right)
 \times (-Y\eee_{x}+X\eee_{y}),
\ENA
where the symbols denote the same as in equation \eqref{eq: B init analytical},
$L_{i}$ are the distances of the twist regions from the mid-plane and
$\phi_{i}$ the two twist angles.
We choose the domain extent and parameters to be the same values as for the
{\em IsoHelix} configuration (\Fig{fig: B_init}, right panel).
Depending on the case we choose either $\phi_{1,2} = \pm\pi/2$ or $\phi_{1,2} = \pm\pi$.
The expected relaxed magnetic field is also of the form \eqref{eq: B relaxed}.
For convenience we will denote this type of initial field as {\em Pontin09}.

\subsection{Diagnostics}

\subsubsection{Force-Freeness}

The final state of our relaxation simulations should be a numerical
approximation to a force-free field.
That is, the final magnetic field
(relaxed state) should approximately satisfy $\nab\times\BB = \alpha\BB$,
where $\alpha$ is constant along magnetic field lines.
In order to quantify the quality of this approximation
we make use of the variable $\alpha^{*}$, defined as
\EQ
\alpha^{*} = \frac{\JJ\cdot\BB}{\BB^{2}}
\EN
\cite{Pontin-Hornig-2009-700-2-ApJ},
where in an exact force-free state $\alpha^{*}$ is constant along field lines.
The magnitude of the variation of $\alpha^{*}$ along a magnetic field line
provides information on the proximity to force-free equilibrium.

In principle one can choose any field line and test how much $\alpha^{*}$
varies,
but that would require the tracing of field lines, which is computationally
expensive and would need high precision.
To circumvent this difficulty we choose the central line interval
\EQ
s_{\alpha} = \left\{(0,0,Z): Z \in [-L_{z}/2, L_{z}/2]\right\}.
\EN
For the two configurations
we know that there is one magnetic field line lying on $s_{\alpha}$ that is
invariant in time (by symmetry).
Therefore, we monitor the maximum difference of $\alpha^{*}$ between any
two points on $s_{\alpha}$ defined as
\EQ \label{eq: epsilon star}
\epsilon^{*} = \max_{i,j}{\left(a_{r}\frac{\alpha^{*}(\XX_{i}) - \alpha^{*}(\XX_{j})}
{|\XX_{i}-\XX_{j}|}\right)};
\quad \XX_{i}, \XX_{j} \in s_{\alpha}.
\EN

\subsubsection{Deviation from the Exact Known Equilibrium}

For the {\em IsoHelix} field
one can simply use the deviation between the exact and numerical results as a measure
of the quality of the final state.
The standard deviation of the magnetic field is simply
\EQ
\sigma_{\BB} = \sqrt{\frac{1}{N}\sum_{ijk} (\BB(\XX_{ijk})-\BB_{\rm relax}(\XX_{ijk}))^{2}},
\EN
with the analytically computed magnetic field $\BB_{\rm relax}(\XX_{ijk})$
for $t \to \infty$ and the total number of grid points $N$.

In analogy, the deviation from the exact grid deformation is
\EQ
\sigma_{\xx} = \sqrt{\frac{1}{N}\sum_{ijk} (\xx(\XX_{ijk})-\xx_{\rm relax}(\XX_{ijk}))^{2}},
\EN
with the analytically computed grid $\xx_{\rm relax}(\XX_{ijk})$ for $t \to \infty$
given by equation \eqref{eq: x analytical relaxed}.

\subsubsection{Convexity}

Certain mimetic methods have been shown to be stable for convex cells
\cite{Hyman-Steinberg-2004-47-1565-CompMathAppl,
Lipnikov-Manzini-2014-257-1163-JCompPhys}.
For concave cells there is no such proof.
It is, therefore, important to monitor the convexity of the cells.
To somewhat simplify the analysis and still retain significance, we define
a convexity parameter associated with grid points, although
convexity is a property of polygons.
At each node $\XX_{ijk}$ one can define eight trihedra composed by its
three nearest neighbors in index space $ijk$.
The three vectors for the trihedra are given as
\EQA
&&{\rm d}\xx^{\lambda} = \xx_{i+\delta_{i},j,k} - \xx_{ijk} \nonumber \\
&&{\rm d}\xx^{\mu}     = \xx_{i,j+\delta_{j},k} - \xx_{ijk} \nonumber \\
&&{\rm d}\xx^{\nu}     = \xx_{i,j,k+\delta_{k}} - \xx_{ijk}; \quad
\delta_{i}, \delta_{j}, \delta_{k} \in \{-1,1\} \label{eq: dx for convexity}
\ENA
and the convexity is defined as
\EQ \label{eq: convexity}
\kappa(\XX_{ijk}) = \left\{
\begin{array}{cc}
1  &
\mathrm{sgn}(\det({\rm d}\xx^{\lambda}{\rm d}\xx^{\mu}{\rm d}\xx^{\nu})) = 
\delta_{i}\delta_{j}\delta_{k} \\
-1 & {\rm otherwise.}
\end{array}
\right.
\EN

\subsubsection{Magnetic Energy}

As discussed above, a force-free magnetic field corresponds to a minimum of
the magnetic energy.
It can be demonstrated that the magneto-frictional evolution equation
\eqref{eq: magneto friction} implies a monotonic decay of the magnetic energy
\cite{Craig-Sneyd-1986-311-451-ApJ,Yang-Sturrock-1986-309-383-ApJ}.
The reliability of the methods applied here and the quality of the relaxation
is consequently also measured by the evolution of the magnetic energy
in the volume $V$
\EQ
\EM=\int_V \BB^2/2\,{\rm d}V.
\EN
Its numerical computation on a moving grid is not trivial, since the volume
${\rm d}V$ surrounding each grid node changes in time.
This volume is given by the determinant $\Delta$ of the Jacobian matrix
multiplied by the corresponding undistorted volume 
${\rm d}X{\rm d}Y{\rm d}Z$.
Boundary points need to be weighted by a factor $\zeta$, as part of their
volume lies outside the domain.
For grid points lying on domain faces $\zeta = 1/2$, on edges $\zeta = 1/4$ and on
corners $\zeta = 1/8$.
Thus, the magnetic energy is 
\EQ \label{eq: magnetic energy}
\EM = \frac{1}{2}\sum_{ijk} \zeta(\XX_{ijk})\ \BB^{2}(\XX_{ijk})\ \Delta(\XX_{ijk})\
{\rm d}X{\rm d}Y{\rm d}Z.
\EN

\section{Quality of the Force-Free Approximation}

Here we describe results obtained using the GLEMuR code with mimetic differential 
operators based on only nearest neighbors, as described in section \ref{sec: nearest}.
These
are compared with results using the classical approach with second-order spatial finite differences.

\subsection{Evolution of Diagnostic Parameters}

\begin{figure*}[t!]\begin{center}
\includegraphics[width=0.49\columnwidth]{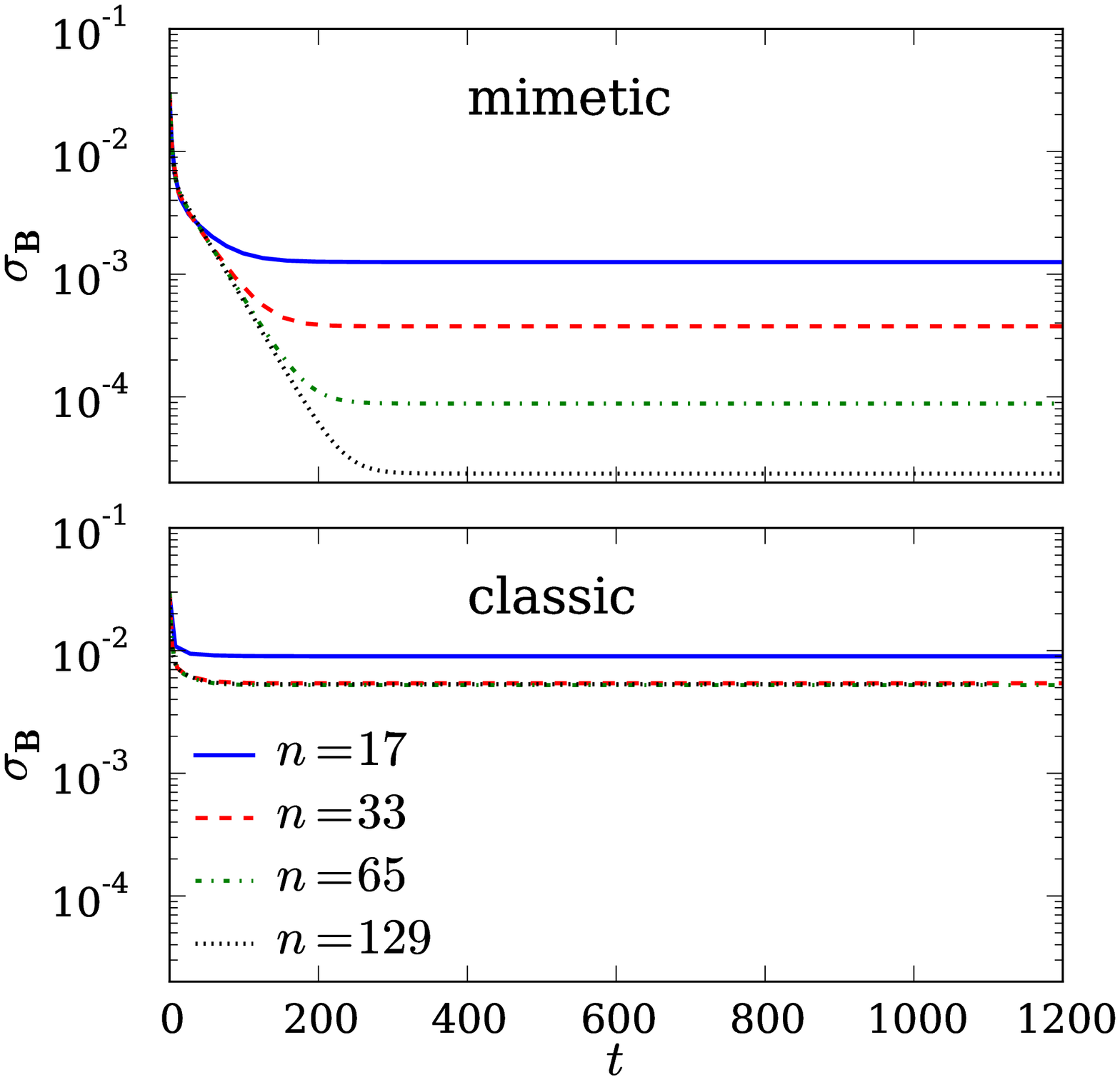}
\includegraphics[width=0.49\columnwidth]{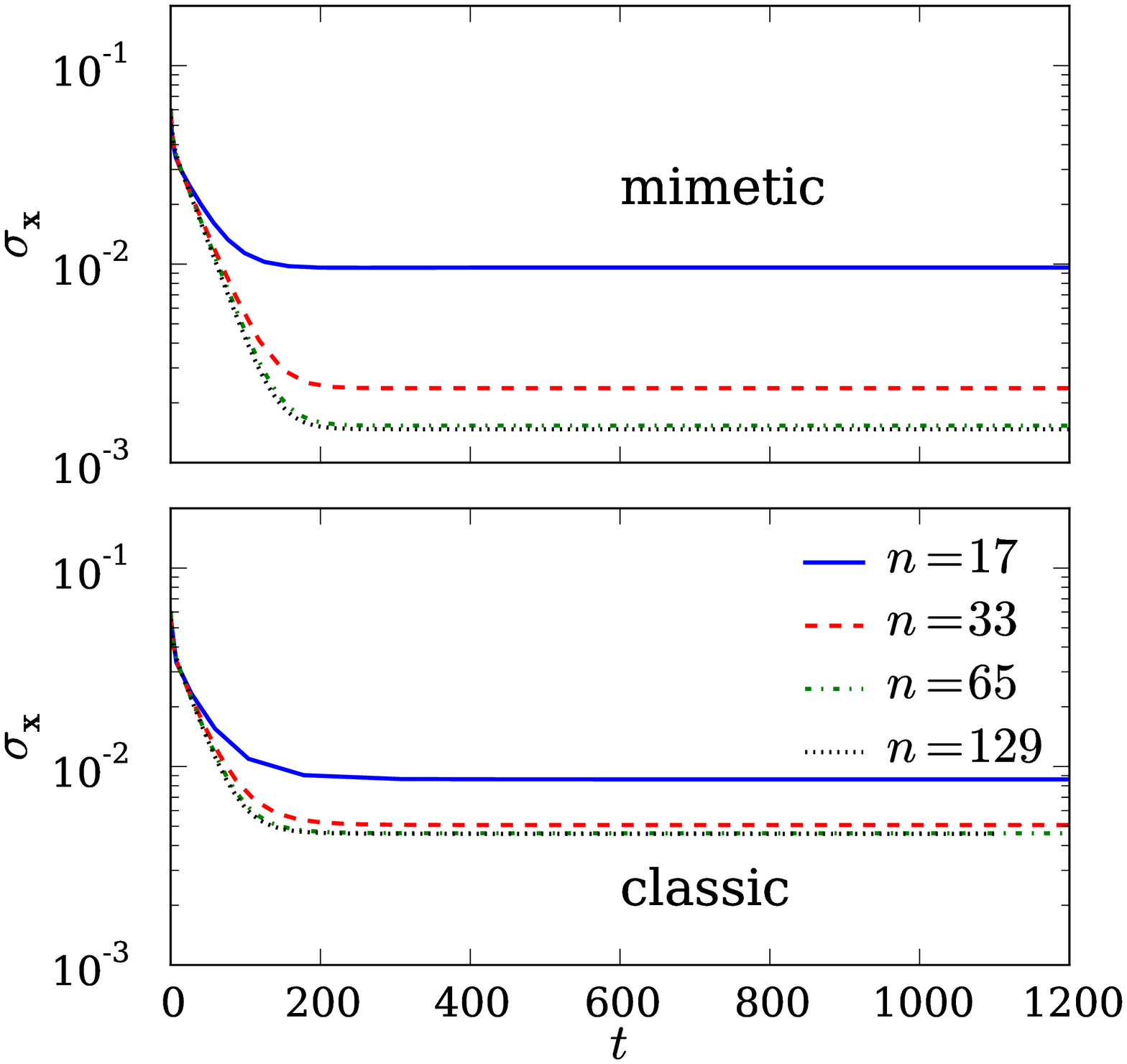} \\
\includegraphics[width=0.49\columnwidth]{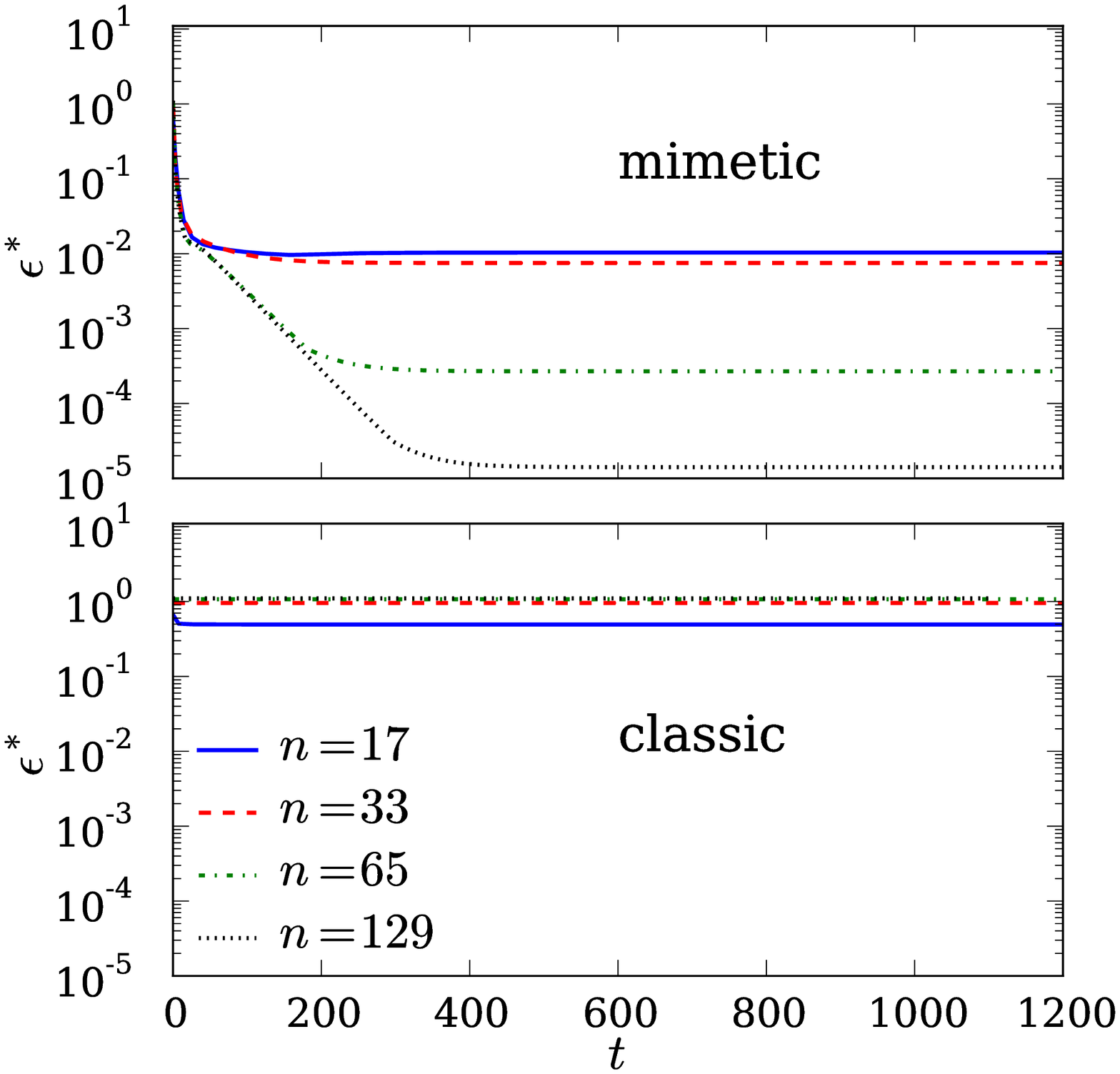}
\includegraphics[width=0.49\columnwidth]{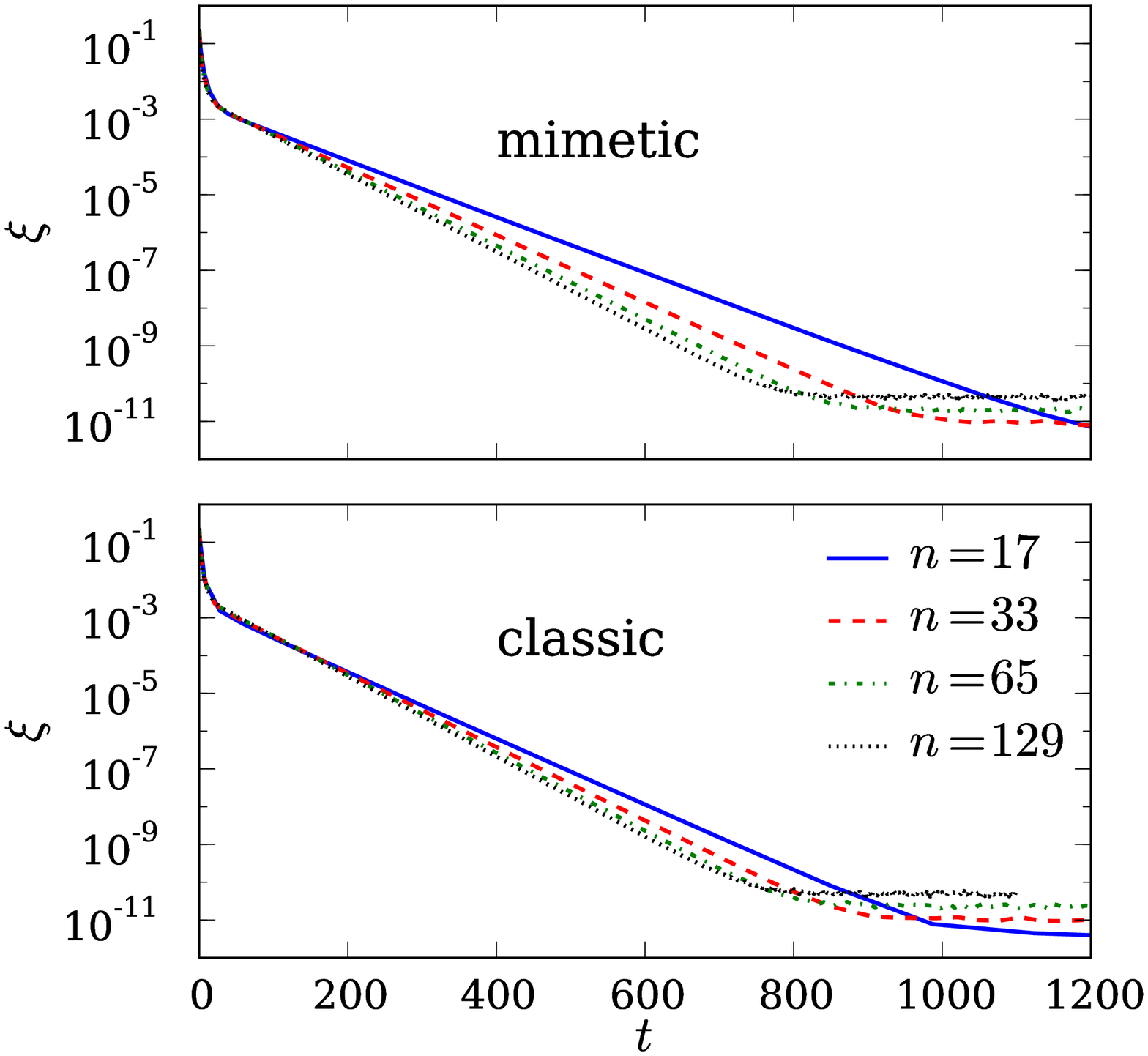} \\
\end{center}\caption[]{
Comparison of the quality of the field relaxation for the mimetic
approach and the classical method as evolution in time for the
{\em IsoHelix} field ($\phi = \pi/4$).
The mimetic approach results in improved quality of the relaxed state as measured
in particular by $\sigma_{\BB}$ and $\sigma_{\xx}$, as well as $\epsilon^{*}$.
}\label{fig: an_pi_4}\end{figure*}

As the magnetic field evolves, it approaches the relaxed state, which is captured
by the decay of the diagnostic variables $\epsilon^{*}$
for the {\em Pontin09} field
and, additionally, $\sigma_{\BB}$ and $\sigma_{\xx}$ for the {\em IsoHelix} field
(\Fig{fig: an_pi_4}, \Tabs{tab: IsoHelix}{tab: Pontin09}).

\begin{table}[b!]
\footnotesize
\caption{
Asymptotic values of the diagnostic parameters as a function of the resolution $n$,
twist angle $\phi$ and method for the numerical derivatives for the {\em IsoHelix}
configuration.
Runs marked by $^{\dagger}$ denote the use of double precision arithmetic (64 bit) in
contrast to single precision (32 bit).
Hyphens mark simulation runs which do not converge.
}
\vspace{12pt}\centerline{\begin{tabular}{llccccc}
\hline \hline
$\phi$ & method & $n$ & $\xi$ & $\epsilon^{*}$ &
$\sigma_{\BB}$ & $\sigma_{\XX}$ \\
\hline
$\pi/4$ & Mimetic$^{\dagger}$ & $17$ & $3.5 {\rm e}{-12}$ & $1.0 {\rm e}{-2}$ &
$1.3 {\rm e}{-3}$ & $9.6 {\rm e}{-3}$ \\
$\pi/4$ & Mimetic$^{\dagger}$ & $33$ & $9.4 {\rm e}{-12}$ & $7.5 {\rm e}{-3}$ &
$3.8 {\rm e}{-4}$ & $2.4 {\rm e}{-3}$ \\
$\pi/4$ & Mimetic$^{\dagger}$ & $65$ & $2.1 {\rm e}{-11}$ & $2.7 {\rm e}{-4}$ &
$8.8 {\rm e}{-5}$ & $1.5 {\rm e}{-3}$ \\
$\pi/4$ & Mimetic$^{\dagger}$ & $129$ & $4.4 {\rm e}{-11}$ & $1.4 {\rm e}{-5}$ &
$2.4 {\rm e}{-5}$ & $1.5 {\rm e}{-3}$ \\
$\pi/2$ & Mimetic & $17$ & $5.0 {\rm e}{-5}$ & $8.3 {\rm e}{-2}$ &
$4.4 {\rm e}{-3}$ & $2.4 {\rm e}{-2}$ \\
$\pi/2$ & Mimetic & $33$ & $3.0 {\rm e}{-5}$ & $6.1 {\rm e}{-2}$ &
$1.5 {\rm e}{-3}$ & $7.1 {\rm e}{-3}$ \\
$\pi/2$ & Mimetic & $65$ & $1.5 {\rm e}{-4}$ & $2.3 {\rm e}{-3}$ &
$3.4 {\rm e}{-4}$ & $5.9 {\rm e}{-3}$ \\
$\pi/2$ & Mimetic & $129$ & $6.1 {\rm e}{-4}$ & $2.0 {\rm e}{-4}$ &
$1.3 {\rm e}{-4}$ & $5.8 {\rm e}{-3}$ \\
$\pi$ & Mimetic & $17$ & -- & -- & -- & -- \\
$\pi$ & Mimetic & $33$ & $3.7 {\rm e}{-4}$ & $5.4 {\rm e}{-1}$ &
$6.4 {\rm e}{-3}$ & $2.7 {\rm e}{-2}$ \\
$\pi$ & Mimetic & $65$ & $1.1 {\rm e}{-3}$ & $2.7 {\rm e}{-2}$ &
$1.4 {\rm e}{-3}$ & $2.1 {\rm e}{-2}$ \\
$\pi$ & Mimetic & $129$ & $4.7 {\rm e}{-3}$ & $2.4 {\rm e}{-3}$ &
$1.0 {\rm e}{-3}$ & $2.3 {\rm e}{-2}$ \\
\hline
$\pi/4$ & Classic$^{\dagger}$ & $17$ & $3.7 {\rm e}{-12}$ & $4.9 {\rm e}{-1}$ &
$9.0 {\rm e}{-3}$ & $8.6 {\rm e}{-3}$ \\
$\pi/4$ & Classic$^{\dagger}$ & $33$ & $1.0 {\rm e}{-11}$ & $9.6 {\rm e}{-1}$ &
$5.4 {\rm e}{-3}$ & $5.1 {\rm e}{-3}$ \\
$\pi/4$ & Classic$^{\dagger}$ & $65$ & $2.3 {\rm e}{-11}$ & $1.1$ &
$5.2 {\rm e}{-3}$ & $4.6 {\rm e}{-3}$ \\
$\pi/4$ & Classic$^{\dagger}$ & $129$ & $5.0 {\rm e}{-11}$ & $1.1$ &
$5.3 {\rm e}{-3}$ & $4.6 {\rm e}{-3}$ \\
$\pi/2$ & Classic & $17$ & $4.4 {\rm e}{-5}$ & $9.9 {\rm e}{-1}$ &
$2.0 {\rm e}{-2}$ & $3.5 {\rm e}{-2}$ \\
$\pi/2$ & Classic & $33$ & $7.3 {\rm e}{-5}$ & $1.9$ &
$1.9 {\rm e}{-2}$ & $2.4 {\rm e}{-2}$ \\
$\pi/2$ & Classic & $65$ & $1.5 {\rm e}{-4}$ & $2.2$ &
$2.0 {\rm e}{-2}$ & $2.0 {\rm e}{-2}$ \\
$\pi/2$ & Classic & $129$ & $7.4 {\rm e}{-4}$ & $2.2$ &
$2.1 {\rm e}{-2}$ & $1.8 {\rm e}{-2}$ \\
$\pi$ & Classic & $17$ & $5.1 {\rm e}{-5}$ & $2.0$ &
$5.5 {\rm e}{-2}$ & $1.3 {\rm e}{-1}$ \\
$\pi$ & Classic & $33$ & $8.1 {\rm e}{-5}$ & $3.8$ &
$6.9 {\rm e}{-2}$ & $1.2 {\rm e}{-1}$ \\
$\pi$ & Classic & $65$ & $3.6 {\rm e}{-4}$ & $4.3$ &
$7.5 {\rm e}{-2}$ & $9.8 {\rm e}{-2}$ \\
$\pi$ & Classic & $129$ & $8.3 {\rm e}{-3}$ &$4.4$ &
$7.6 {\rm e}{-2}$ & $7.4 {\rm e}{-2}$ \\
\hline \hline
\end{tabular}}
\label{tab: IsoHelix}
\end{table}

\begin{table}[b!]
\footnotesize
\caption{
Asymptotic values of the diagnostic parameters as a function of the resolution $n$
and method for the numerical derivatives for the {\em Pontin09}
configuration ($\phi = \pi/2$).
}
\vspace{12pt}\centerline{\begin{tabular}{lccc}
\hline \hline
method & $n$ & $\xi$ & $\epsilon^{*}$ \\
\hline
Mimetic & $17$ & $5.4 {\rm e}{-5}$ & $1.1 {\rm e}{-1}$ \\
Mimetic & $33$ & $3.3 {\rm e}{-4}$ & $1.8 {\rm e}{-2}$ \\
Mimetic & $65$ & $8.9 {\rm e}{-4}$ & $7.4 {\rm e}{-4}$ \\
Mimetic & $129$ & $4.0 {\rm e}{-3}$ & $6.4 {\rm e}{-4}$ \\
\hline
Classic & $17$ & $1.5 {\rm e}{-4}$ & $5.0 {\rm e}{-1}$ \\
Classic & $33$ & $1.8 {\rm e}{-4}$ & $7.9 {\rm e}{-1}$ \\
Classic & $65$ & $9.8 {\rm e}{-4}$ & $8.5 {\rm e}{-1}$ \\
Classic & $129$ & $2.9 {\rm e}{-3}$ & $8.6 {\rm e}{-1}$ \\
\hline \hline
\end{tabular}}
\label{tab: Pontin09}
\end{table}

The evolution of $\epsilon^{*}$ provides one window into the quality of
the force-free field obtained.
Comparing the results for the mimetic and
classical approaches, we find that for all of the configurations
investigated here
(\Fig{fig: an_pi_4}, \Tabs{tab: IsoHelix}{tab: Pontin09})
the mimetic approach gives a greatly improved relaxation as measured by $\epsilon^{*}$.
The classical method converges to values of the order of one, almost independently
of the resolution, while the mimetic approach improves this by more than four
orders of magnitude with convergence towards higher resolutions
(\Tabs{tab: IsoHelix}{tab: Pontin09}).

In addition to the above, one can also monitor directly the normalized
maximum of the Lorentz force in the domain
\EQ
\xi = \max{\frac{|\JJ\times\BB|}{\BB^{2}}}.
\EN
For both methods this can be seen to decay to extremely small values
(\Fig{fig: an_pi_4}, \Tabs{tab: IsoHelix}{tab: Pontin09}) that
are essentially limited only by numerical roundoff errors.
However, as was shown by Pontin et al.\ \cite{Pontin-Hornig-2009-700-2-ApJ}, these numbers
can be highly misleading.
In particular, it was shown that for the classical method the Lorentz force
is minimized at the expense of the accuracy of, in particular, $\nab\times\BB$.
Comparing plots for both the classical and mimetic methods, we see that
$\xi$ continues to decrease even after all independent measures of the
force-freeness stabilize to a constant level.
As a result, we do not consider the directly calculated value of $\xi$ to be
a reliable measure of the true accuracy of the force-free approximation.

Further, the value of $\xi$ strongly depends on the resolution and the tolerance
$\Lambda_{0}$.
The former can even have a negative effect if $\Lambda_{0}$ is chosen to be the
same irrespective of the resolution.
We explain this by the error of the grid deformation during the time stepping
(Eq.~\ref{eq: time stepping}), where $\Lambda_{0}$ is set to similar
values for different resolutions.
If the grid error is the same for high and low resolutions, the error in the
derivatives is higher for smaller grid separations, which is why we see higher
values for $\xi$.

\subsection{Deviations from the Analytical Solution}
For the \emph{IsoHelix} configuration, we can directly assess the accuracy of the method
by comparing the magnetic field and grid distortion with the known exact values as measured
by $\sigma_{\BB}$ and $\sigma_{\xx}$.
Like $\xi$ and $\epsilon^{*}$, $\sigma_{\BB}$ and $\sigma_{\xx}$ decrease over time,
indicating the relaxation of the field towards a force-free state (\Fig{fig: an_pi_4}).
For the mimetic approach there is a reduction in these quantities, in some cases by more than 
two orders of magnitude.
We also confirm strong improvements with increasing resolution.
Their monotonic decay serves us as reassurance that the mimetic approach is very
well suited for studying relaxation processes.

\subsection{Magnetic Energy}

\begin{figure}[t!]\begin{center}
\includegraphics[width=.49\columnwidth]{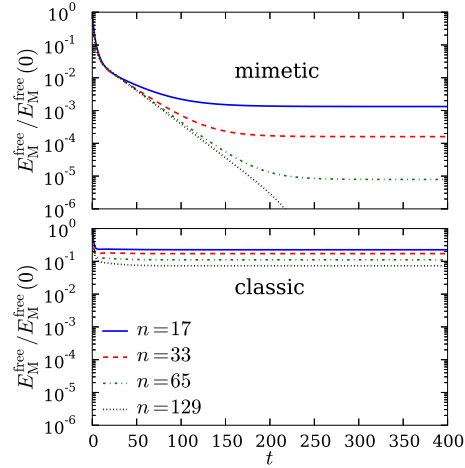}
\end{center}\caption[]{
Normalized relative free magnetic energy in time for the {\em IsoHelix} configuration
with $\phi = \pi/4$ using the mimetic approach (upper panel)
and classical approach (lower panel).
As expected, the energy decreases monotonically.
}\label{fig: pontin_pi_2_EMfree}\end{figure}

Motivated by previous predictions on the magnetic energy evolution
\cite{Craig-Sneyd-1986-311-451-ApJ,Yang-Sturrock-1986-309-383-ApJ} we monitor the 
free magnetic energy $\EM^{\rm free} = \EM - \EM^{0}$, where $\EM^{0}$ is the
magnetic energy stored in the homogeneous background field
$\BB_{\rm bkg} = B_{0}\eee_{z}$.
From that analysis we confirm that $\EM^{\rm free}$ and $\EM$ decrease
monotonically in time (\Fig{fig: pontin_pi_2_EMfree}),
which is well established even for very low grid resolutions.
The classical approach allows the energy to decay only down to a certain
threshold while the mimetic approach leads to the expected decay of the
free energy.
This behavior also serves as additional verification that all applied methods
are able to reproduce correct results within their limits.

\subsection{Grid Convexity and Stability}

Relaxation of the magnetic fields used here results in an untwisted magnetic
field, which is achieved by twisting the grid in the opposite sense to the initial magnetic field
twist.
Increasing the field's initial twist ($\phi$) also increases the expected grid
distortion of the relaxed state, as the field unwinds itself.
Such high distortions lead to concave grid cells, particularly for low resolutions, for which the
mimetic operators might not yield a good approximation
\cite{Lipnikov-Manzini-2014-257-1163-JCompPhys}.

\begin{figure*}[t!]\begin{center}
\includegraphics[width=.32\columnwidth]
{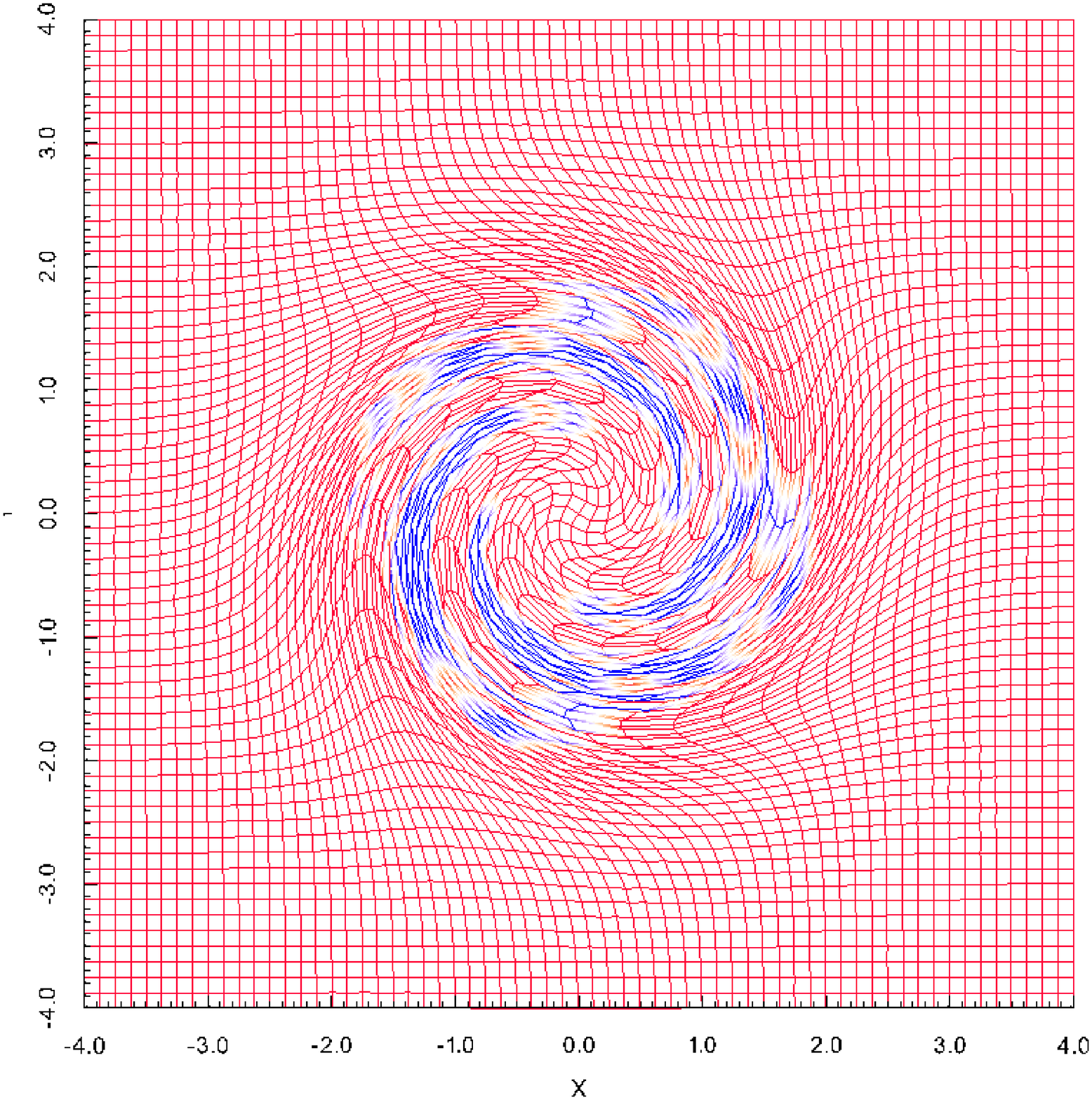}
\includegraphics[width=.32\columnwidth]
{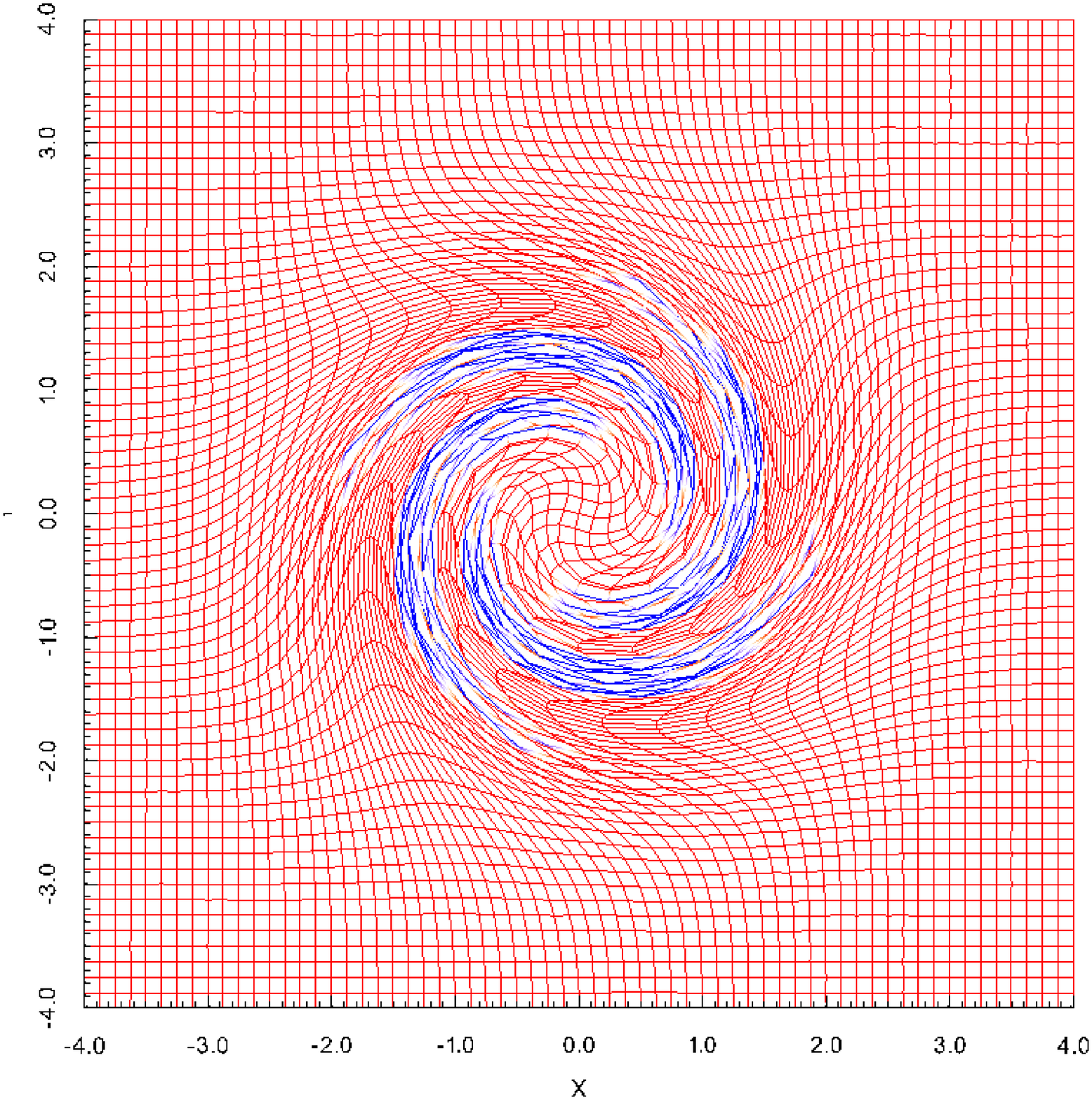}
\includegraphics[width=.32\columnwidth]
{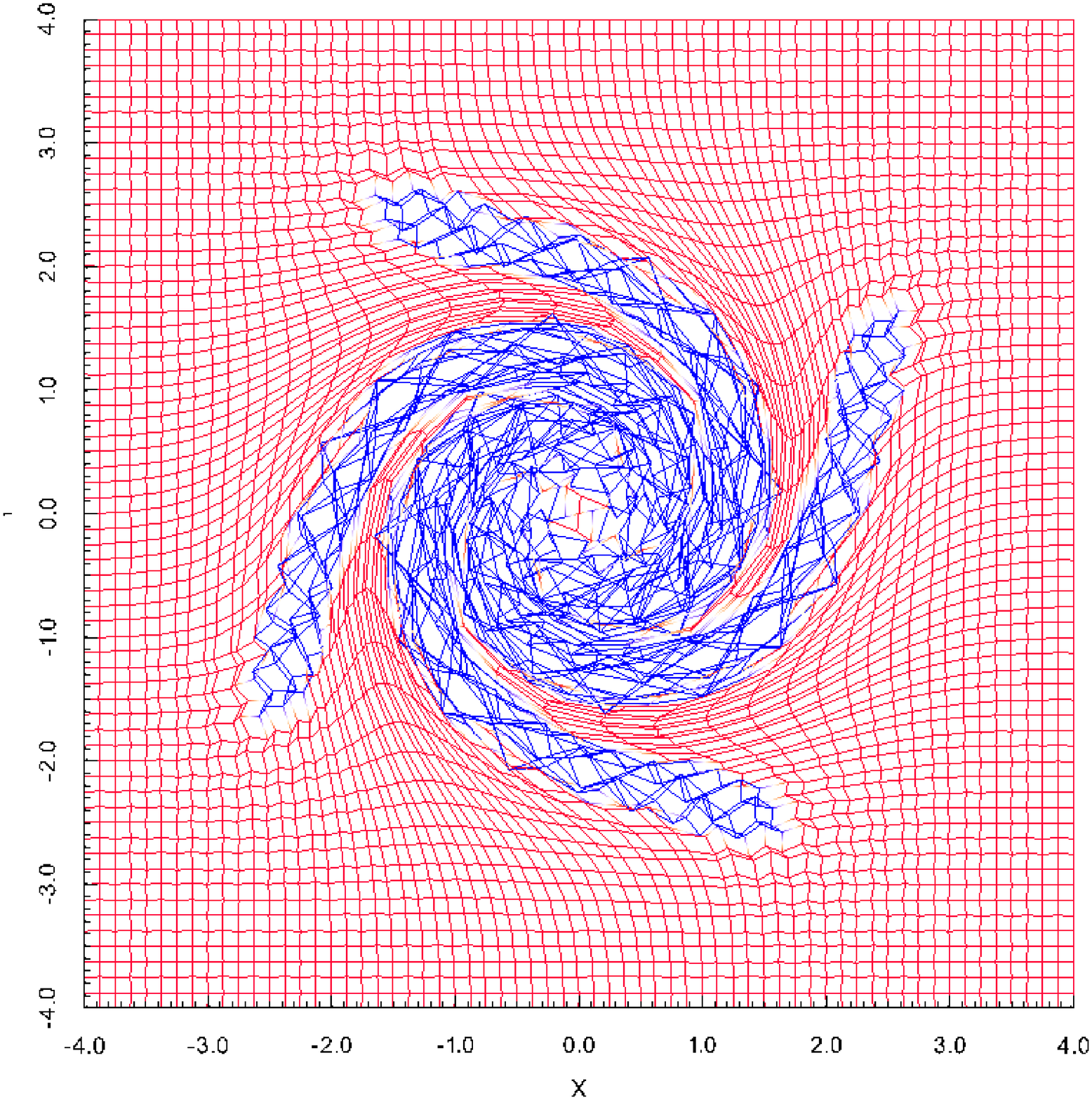}
\end{center}\caption[]{
Grid distortion as seen in the $ij$-index plane at $Z = 0$ for the final stage
of the simulation for the {\em Pontin09} configuration ($\phi = \pi$).
Red grid nodes denote convexity, while blue denote concavity
as defined in equation \eqref{eq: convexity}.
The left panel shows the grid close to relaxation using the classical method.
Second and third panel show the same configuration using the
mimetic method at time $t = 200$ and $t = 238.2$, respectively.
The mimetic method breaks down shortly after the grid becomes locally concave.
}\label{fig: pontin09 grid}\end{figure*}

\begin{figure}[t!]\begin{center}
\includegraphics[width=.49\columnwidth]{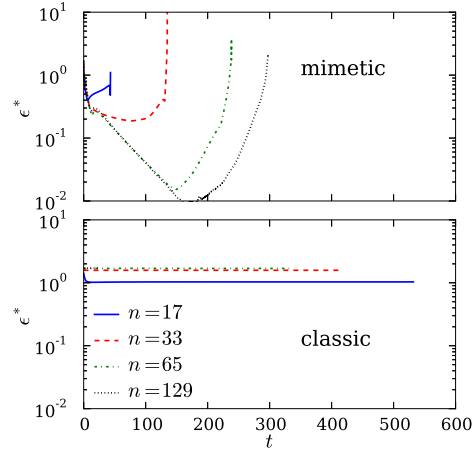}
\end{center}\caption[]{
Time evolution for the {\em Pontin09} configuration ($\phi_{1/2} = \pm\pi$)
of $\epsilon^{*}$ for $\JJ$ computed by using the mimetic
approach (upper panel) and classical derivatives (lower panel).
For $\epsilon^{*}$, the mimetic method is far superior in creating
a force-free field but lacks in stability for this particular field.
}\label{fig: pontin_pi}\end{figure}

The grid distortion is clearly seen in \Fig{fig: pontin09 grid} where we plot the grid at
the mid-plane $Z = 0$ at an intermediate time for the {\em Pontin09} configuration
($\phi = \pi$).
We also plot the convexity, as defined in equation \eqref{eq: convexity},
where red represents convexity and blue concavity.
Applying the classical  method we find that the grid becomes
locally concave (\Fig{fig: pontin09 grid}, left panel) but the simulation
remains stable.
The mimetic approach also leads to concave cells
(\Fig{fig: pontin09 grid}, central panel) which subsequently causes jagged
grid distortions and the
method breaks down (\Fig{fig: pontin09 grid}, right panel).
At this time, we see a blow up of the diagnostic parameters
(\Fig{fig: pontin_pi}) together with a
drop of the time step by several orders of magnitude, at which point
the simulation is stopped.
Increased grid resolution can delay this blow up.
Moreover, it should be stressed that while the classical approach is stable in
this case, it does not result
in an improved relaxed state, as measured by $\epsilon^{*}$. 
Indeed, the mimetic approach before the blow up
provides by orders of magnitude a better force-free approximation, 
see Figure \ref{fig: pontin_pi}.

\subsection{Next-Nearest-Neighbors Mimetic Approach}

\begin{figure}[t!]\begin{center}
\includegraphics[width=.49\columnwidth]{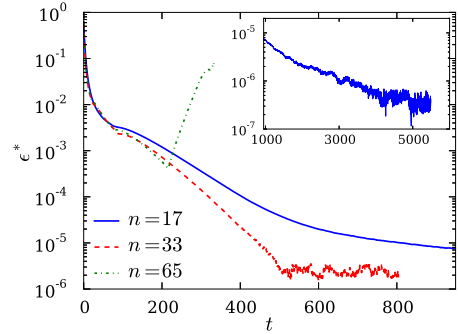}
\end{center}\caption[]{
Time evolution of the force-free measure $\epsilon^{*}$ for the
next-nearest-neighbor mimetic approach using the {\em IsoHelix} configuration
with $\phi = \pi/4$.
The inset shows the time evolution for $n =17$ for longer times.
Compared to the nearest neighbor method there is an improvement of 5 orders of
magnitude as measured by $\epsilon^{*}$.
}\label{fig: an_pi_4_stokes4th}\end{figure}

Here we apply our next-nearest-neighbor curl operator to compute $\JJ = \nab\times\BB$,
 described in Section \ref{nextnearestsec}.
Subject to this study is the field for which we know its analytical solution
(Eq.~\eqref{eq: B init analytical}) with $\phi = \pi/4$.
For the evolution of $\xi$, $\sigma_{\BB}$ and $\sigma_{\xx}$ we observe almost
identical behavior as for the nearest neighbor approach.
In that respect there is no advantage of this method over the nearest
neighbor method.
By contrast, for $\epsilon^{*}$ we observe an improvement of up to 5 orders
of magnitude (\Fig{fig: an_pi_4_stokes4th}).
However, this method proves to be unstable for all other configurations discussed herein.
Indeed, the numerical instability sets in even before the grid becomes
concave, which severely limits its applicability. 
This suggests that including additional grid points in the mimetic approach is 
in general not likely to be fruitful.

\section{Performance}

We compare the computation time for $\JJ=\nab\times\BB$ for the classical
direct approach, as used by Craig et al.\ \cite{Craig-Sneyd-1986-311-451-ApJ}, with the
mimetic approach.
Since the simulation is performed on an Nvidia graphics card model
GTX 765M, we use the
{\em NVIDIA Visual Profiler} tool to compare the computation time of the
computation kernels for a resolution of $65^{3}$ grid points.
For computing $\JJ$
the classical approach requires a typical time of about $37.56 {\rm ms}$,
while the mimetic approach only needs $19.82 {\rm ms}$.

Summing up all computationally intensive floating point operations,
like multiplications, divisions and roots, we know that there are $462$
multiplications and $7$ divisions for the classical method.
For the mimetic approach there are only $156$ multiplications and one division,
but $12$ roots.
In both cases, multiplications and divisions by a factor of $2^{n}$ with
$n \in \mathbb{N}$ are excluded from the operation count,
since they only require a bitwise shift.
The difference in computational working load approximately reflects
the measured timings.

Currently our code runs on single GPUs only.
This means that running simulations on multiple graphics cards, like on a cluster,
would not increase the computational speed.
Since efficient multi-GPU computation for finite difference schemes is rather
labor intensive to design we left this open for future work.
However, the code is currently designed such that it can in principle run on
hardware with any number of multiprocessors, and has run on high-end cards like
the Nvidia Tesla K40.
As the development of graphics cards is rapid we will soon be able to use our
code on future hardware without computational penalties.

\section{Conclusions}

The question as to whether for an arbitrary given magnetic field
a corresponding force-free field (Beltrami field) with the same topology
exists, and if so whether it is smooth,
is an important unsolved problem in plasma physics.
We have  presented here a new code that performs a relaxation of a magnetic field towards
a force-free state using a Lagrangian numerical scheme.
The method strictly preserves the magnetic flux and the topology of magnetic
field lines.
In contrast to other implementations we use mimetic operators for the
spatial derivatives in order to improve accuracy for high distortions
of the grid. 
We implement the scheme in a code which runs on graphical processing units
(GPU), which leads to an enhanced computing speed compared to previous
relaxation codes.
Compared with schemes using direct derivatives we find 
that the final state of the simulation approximates a
force-free magnetic field with a significantly higher accuracy. 
Furthermore, as expected, this accuracy improves as the resolution increases.
It is found, however, that the method is only numerically stable so long as the cells
of the numerical grid remain convex.
This places a restriction on the proximity of the 
initially prescribed field to the corresponding force-free field.
Increasing the number of points 
used in the scheme to consider next-nearest-neighbors is found to strongly compromise the
stability, indicating that this is not a fruitful approach for such schemes.

\appendix

\section{Derivation of Eqs.~\eqref{eq: pull back} and \eqref{eq: B transformation}}
\label{sec: induction lie}

To extend the initial discussion about the ideal evolution
we express Eq.~\eqref{eq: ideal induction} in terms of a
Lie-derivative of a differential 2-form $\beta$ associated with the vector $\BB$.
The relation between the 2-form and the vector $\BB$ is given by the interior
product $\beta = i_{\BB}\mu$ where $\mu$ is the standard volume form in the domain.
In Cartesian coordinates $(X^1,X^2,X^3)$ this reads
\[ \beta = \beta_{23} \dd X^2 \wedge \dd X^3 +\beta_{13} \dd X^1 \wedge \dd X^3 + \beta_{12} \dd X^1 \wedge \dd X^2, \]
where 
\[ \beta_{12} = B_3, \ \beta_{23} = B_1, \ \beta_{13} = - B_2. \]
Hence, Eq.~\eqref{eq: ideal induction} is equivalent to
\[ \frac{\partial}{\partial t} \beta (\XX,t) +  {\cal L}_{\uu} \beta (\XX,t) = 0, \]
where ${\cal L}_{\uu}$ is the Lie-derivative with respect to $\uu$.
This is in turn the differential formulation of 
\[ (\xx^{*}(t)\beta)(\XX,t)  = \beta(\XX,0), \] 
where the star indicates the pull-back operation (see
\cite[pp.~370]{abraham1988manifolds} and \cite[pp.~140-3]{frankel2011geometry}).
Writing this out we get
\begin{eqnarray*}
(\xx^{*}(t) \beta)(\XX,t) & = & \beta_{i j}(\xx(\XX,t),t)
\frac{\partial x^{i}}{\partial X^{k}}
\frac{\partial x^{j}}{\partial X^{l}} \dd X^{k}\wedge \dd X^{l},  \
\quad i,j, k, l \in \{1,2,3\} \ \mbox{and} \  i < j.
\end{eqnarray*}
One can solve this equation for $\beta(\XX,t)$, using the formula for the adjoint
of the Jacobian matrix.
Translating this back into components of the vector field $\BB$  leads to equation
\eqref{eq: B transformation}.

\bibliographystyle{siam}
\bibliography{references}

\begin{thebibliography}{10}

\bibitem{abraham1988manifolds}
{\sc R.~Abraham, J.E. Marsden, and T.S. Ratiu}, {\em Manifolds, Tensor
  Analysis, and Applications}, no.~v. 75 in Applied Mathematical Sciences,
  Springer New York, 1988.

\bibitem{alfven1943}
{\sc H.~{Alfv{\'e}n}}, {\em {On the Existence of Electromagnetic-Hydrodynamic
  Waves}}, Arkiv for Astronomi, 29 (1943), pp.~1--7.

\bibitem{BatchelorFrozeIn1950RSPSA}
{\sc G.~K. {Batchelor}}, {\em {On the Spontaneous Magnetic Field in a
  Conducting Liquid in Turbulent Motion}}, Proc. R. Soc. Lond. A., 201 (1950),
  pp.~405--416.

\bibitem{Cash-Karp-1990-16-3-ACMTransMathSoft}
{\sc J.~R. Cash and A.~H. Karp}, {\em A variable order {Runge-Kutta} method for
  initial value problems with rapidly varying right-hand sides}, ACM T. Math.
  Software, 16 (1990), pp.~201--222.

\bibitem{Chodura1981}
{\sc R.~Chodura and A.~Schl{\" u}ter}, {\em {A 3D code for MHD equilibrium and
  stability}}, J. Comput. Phys., 41 (1981), p.~68.

\bibitem{Craig-Sneyd-1986-311-451-ApJ}
{\sc I.~J.~D. Craig and A.~D. Sneyd}, {\em {A dynamic relaxation technique for
  determining the structure and stability of coronal magnetic fields}},
  Astrophys. J., 311 (1986), pp.~451--459.

\bibitem{Craig2005}
\leavevmode\vrule height 2pt depth -1.6pt width 23pt, {\em {The Parker Problem
  and the Theory of Coronal Heating}}, Solar Physics, 232 (2005), p.~41.

\bibitem{frankel2011geometry}
{\sc T.~Frankel}, {\em The Geometry of Physics: An Introduction}, Cambridge
  University Press, 2011.

\bibitem{Hyman-Shashkov-1997-33-4-CompAthApp}
{\sc J.~M. Hyman and M.~Shashkov}, {\em Natural discretizations for the
  divergence, gradient, and curl on logically rectangular grids}, Comput. Math.
  Appl., 33 (1997), pp.~81--104.

\bibitem{Hyman-Shashkov-1999-151-2-JCompPh}
\leavevmode\vrule height 2pt depth -1.6pt width 23pt, {\em Mimetic
  discretizations for maxwell's equations}, J. Comput. Phys., 151 (1999),
  pp.~881--909.

\bibitem{Hyman-Steinberg-2004-47-1565-CompMathAppl}
{\sc J.~M. Hyman and S.~Steinberg}, {\em The convergence of mimetic
  discretization for rough grids}, Comput. Math. Appl., 47 (2004),
  pp.~1565--1610.

\bibitem{Lipnikov-Manzini-2014-257-1163-JCompPhys}
{\sc K.~Lipnikov, G.~Manzini, and M.~Shashkov}, {\em Mimetic finite difference
  method}, J. Comput. Phys., 257, Part B (2014), pp.~1163--1227.

\bibitem{longcope1994}
{\sc D.~W. {Longcope} and H.~R. {Strauss}}, {\em {The form of ideal current
  layers in line-tied magnetic fields}}, Astrophys.~J., 437 (1994),
  pp.~851--859.

\bibitem{low2010apj}
{\sc B.~C. Low}, {\em {The Parker Magnetostatic Theorem}}, Astrophys.~J., 718
  (2010), pp.~717--723.

\bibitem{Low-2013-768-7-ApJ}
\leavevmode\vrule height 2pt depth -1.6pt width 23pt, {\em Newtonian and
  non-newtonian magnetic-field relaxations in solar-coronal mhd}, Astrophys.
  J., 768 (2013), p.~7.

\bibitem{MoffattBook1978}
{\sc H.~K. Moffatt}, {\em {Magnetic field generation in electrically conducting
  fluids}}, Camb. Univ. Press, 1978.

\bibitem{Moffatt1985}
{\sc H.~K. Moffatt}, {\em {Magnetostatic equilibria and analogous Euler flows
  of arbitrarily complex topology. I - Fundamentals}}, J. Fluid Mech., 159
  (1985), pp.~359--378.

\bibitem{Nickolls-Buck-2008-6-2-Queue}
{\sc J.~Nickolls, I.~Buck, M.~Garland, and K.~Skadron}, {\em Scalable parallel
  programming with cuda}, Queue, 6 (2008), pp.~40--53.

\bibitem{Parker72}
{\sc E.~N. Parker}, {\em {Topological Dissipation and the Small-Scale Fields in
  Turbulent Gases}}, Astrophys. J., 174 (1972), p.~499.

\bibitem{Pontin-Hornig-2009-700-2-ApJ}
{\sc D.~I. Pontin, G.~Hornig, A.~L. Wilmot-Smith, and I.~J.~D. Craig}, {\em
  Lagrangian relaxation schemes for calculating force-free magnetic fields, and
  their limitations}, Astrophys. J., 700 (2009), p.~1449.

\bibitem{NumericalRecipes3}
{\sc W.~H. Press, S.~A. Teukolsky, W.~T. Vetterling, and B.~P. Flannery}, {\em
  {Numerical Recipes 3rd Edition: The Art of Scientific Computing}}, Cambridge
  University Press, 3~ed., 2007.

\bibitem{PriestReconnection2000}
{\sc E.~R. {Priest} and T.~G. {Forbes}}, {\em Magnetic reconnection: MHD theory
  and applications}, 2000.

\bibitem{rapazzo2013}
{\sc A.~F. {Rappazzo} and E.~N. {Parker}}, {\em {Current Sheets Formation in
  Tangled Coronal Magnetic Fields}}, Astrophys.~J.~Lett., 773 (2013), p.~L2.

\bibitem{Syrovatskii71}
{\sc S.~I. Syrovatskii}, {\em {Formation of Current Sheets in a Plasma with a
  Frozen-in Strong Magnetic Field}}, Soviet Journal of Experimental and
  Theoretical Physics, 33 (1971), p.~933.

\bibitem{vanballegooijen1985}
{\sc A.~A. {van Ballegooijen}}, {\em Electric currents in the solar corona and
  the existence of magnetostatic equilibrium}, Astrophys.~J., 298 (1985),
  p.~421.

\bibitem{wilmotsmith2009a}
{\sc A.~L. Wilmot-Smith, G.~Hornig, and D.~I. Pontin}, {\em Magnetic braiding
  and parallel electric fields}, Astrophys. J., 696 (2009), pp.~1339--1347.

\bibitem{Yang-Sturrock-1986-309-383-ApJ}
{\sc W.~H. {Yang}, P.~A. {Sturrock}, and S.~K. {Antiochos}}, {\em {Force-free
  magnetic fields - The magneto-frictional method}}, Astrophys. J., 309 (1986),
  pp.~383--391.

\end{thebibliography}

\end{document}